\documentclass[useAMS,usenatbib,usegraphicx]{mn2e}

\title[The magnetic field in the jet of 3C~346]{The magnetic field and geometry of the oblique shock in the jet of 3C~346}
\author[F. Dulwich et al.]{F.~Dulwich$^{1}$, D.M.~Worrall$^{1}$, M.~Birkinshaw$^{1}$, C.A.~Padgett$^{2}$, E.S.~Perlman$^{3}$ \\
    $^{1}$ H.H. Wills Physics Laboratory, University of Bristol, Tyndall Avenue, Bristol BS8 1TL, UK \\
    $^{2}$ Joint Center for Astrophysics, University of Maryland-Baltimore County, 1000 Hilltop Circle, Baltimore, MD 21250, USA \\
    $^{3}$ Dept. of Physics and Space Sciences, Florida Institute of Technology, 150 West University Boulevard, Melbourne, FL 32901, USA}

\usepackage[totalwidth=480pt, totalheight=680pt]{geometry}
\usepackage{rotating}

\def\sgn{\mathop{\rm sgn}}
\newcommand{\vect}[1]{\vec{#1}}

\begin{document}

\date{Received **}

\pagerange{\pageref{firstpage}--\pageref{lastpage}} \pubyear{2009}

\maketitle

\label{firstpage}

\begin{abstract}
We investigate the brightest regions of the kpc-scale jet in the powerful radio galaxy 3C~346, using new optical {\it Hubble Space Telescope} ({\it HST}) ACS/F606W polarimetry together with {\it Chandra} X-ray data and 14.9~GHz and 22.5~GHz {\it VLA} radio polarimetry. The jet shows a close correspondence in optical and radio morphology, while the X-ray emission shows an $0.80 \pm 0.17$ kpc offset from the optical and radio peak positions. Optical and radio polarimetry show the same apparent magnetic field position angle and fractional polarization at the brightest knot, where the jet undergoes a large kink of almost $70$ degrees in the optical and radio images. The apparent field direction here is well-aligned with the new jet direction, as predicted by earlier work that suggested the kink was the result of an oblique shock. We have explored models of the polarization from oblique shocks to understand the geometry of the 3C~346 jet, and find that the upstream flow is likely to be highly relativistic ($\beta_u = 0.91_{-0.07}^{+0.05}$), where the plane of the shock front is inclined at an angle of $\eta = 51 \pm 11$ degrees to the upstream flow which is at an angle $\theta = 14_{-7}^{+8}$ degrees to our line of sight. The actual deflection angle of the jet in this case is only $22$ degrees.
\end{abstract}


\section{Introduction}\label{sec:intro}

Radio observations of kpc-scale jets in active galaxies have shown \citep[e.g.][]{BP84} that the emissions tend to be highly polarized and exhibit a characteristic synchrotron spectrum. Observations of nearby sources at shorter wavelengths often show a close correspondence to the radio morphology and characteristic spectral-energy distributions, suggesting that synchrotron radiation from energetic particles also dominates the emission in the optical \citep[e.g.][]{Keel88,Fedorenko96} and X-ray \citep*{Worrall01b,Hardcastle01}. High-energy electrons ($\gamma = E/mc^2$ $\sim 10^7$ -- $10^8$) radiate at optical and X-ray wavelengths and have very short synchrotron-emitting lifetimes ($10^2$ -- $10^3$ years in a magnetic field of a few nT), and must therefore be reaccelerated locally to maintain the observed emissions. Lower-energy electrons have much longer lifetimes ($\sim 2 \times 10^6$ years, for $\gamma = 10^4$ and $B = 3$~nT) and are detected primarily at radio wavelengths.

Polarization studies of the synchrotron emission from kpc-scale jets provide valuable information about their structures and magnetic configurations: the observed polarization direction gives an emission-weighted measure of the apparent magnetic field direction along the line of sight, while the fractional polarization indicates the ordering of the field. Multi-band imaging and polarimetry are therefore powerful tools in the study of jets, allowing us to use the emission across the electromagnetic spectrum to probe populations of particles at different energies.

The powerful FR~II radio galaxy 3C~346 lies at redshift $z = 0.16$ and has a 17$^{\rm th}$-magnitude elliptical host at optical wavelengths \citep*[e.g.][]{Laing83}. Although originally classified as an FR~I galaxy by \citet{Laing83}, radio maps of the source show a highly luminous core surrounded by diffuse lobe emission and a highly-distorted eastern jet exhibiting a bright knot structure \citep{Spencer91,Cotton95}: \citet{Spencer91} argued that these features could be best explained by an FR~II radio galaxy being foreshortened by relativistic beaming at a small angle to the line-of-sight, and \citet{Cotton95} used the radio-core dominance and jet/counter-jet ratio to deduce an angle to the line-of-sight $\theta < 32$ degrees and jet speed $\beta > 0.8$. The jet was first detected in the optical using ground-based observations made by \citet{Dey94}. Subsequent {\it Hubble Space Telescope} ({\it HST}) observations showed striking similarities between the optical and radio jet morphologies \citep{deKoff96,deVries97}.

\citet{Worrall05} reported the first detection of X-ray radiation from the brightest knot in the jet of 3C~346 using high-resolution images from the {\it Chandra} X-ray Observatory, finding that the peak of the X-ray emission does not exactly match the peak of the radio and optical emission. The available multiwavelength data allowed \citet{Worrall05} to conclude that the X-ray emission process is synchrotron and to discuss models for the jet. They suggested that the bright knot and accompanying sharp kink in the jet were the result of an oblique shock, and predicted that this should generate a relatively simple polarization pattern with the magnetic field direction at position angle $\sim 20$ degrees. This is consistent with the high (17 per cent) radio polarization measured by \citet{Akujor95}, but sufficiently good high resolution optical and radio polarimetry were not available to probe the magnetic field structure until now. The polarization data allow us to test the prediction that the magnetic field vectors should be largely aligned with the direction of the putative oblique shock.

In this paper we re-examine the 3C~346 jet, using new {\it HST} optical polarimetry \citep{Perlman06} in conjunction with previously unpublished 14.9~GHz and 22.5~GHz {\it VLA} radio polarimetry, and the {\it Chandra} X-ray data. This work probes the jet in sufficient detail for us to test the predictions of \citet{Worrall05} and distinguish between different shock models in 3C~346. We adopt models for oblique and conical shocks from \citet{Cawthorne06}, and use these to explore the geometry of the shock and the upstream flow speed. Section~\ref{sec:observations} describes the multi-wavelength observations we used. We present our results in Section~\ref{sec:results}, and in Section~\ref{sec:models} we examine models of the magnetic field at the shock. Throughout, we adopt values for the cosmological parameters $H_0=70$ km s$^{-1}$ Mpc$^{-1}$, $\Omega_{{\Lambda}0}=0.7$ and $\Omega_{m0}=0.3$. Spectral indices are defined as the slope of the power-law relating the flux $S_\nu$ at a frequency $\nu$ according to $S_\nu \propto \nu^{-\alpha}$. At the redshift of 3C~346, one arcsecond corresponds to a projected distance of 2.76 kpc.

\section{Observations}\label{sec:observations}
\subsection{Optical data}\label{subsec:opticaldata}
All observation details are summarised in Table~\ref{tab:obs}. Optical {\it HST} images of 3C~346 were obtained on 2003 August 19 with the ACS/WFC instrument using the wide-band F606W filter and the three optical polarizers, each in a 502 second exposure. Optical polarimetry on 3C~346 was first presented by \citet{Perlman06}. Data were flat-fielded and bias-corrected using standard packages in {\sc iraf/stsdas}, and {\sc crreject} was run to remove cosmic-ray events \citep[see][for further details]{Perlman06,Perlman99}. The ACS images were combined using multidrizzle \citep{Koekemoer02}, allowing for corrections in the chip geometry \citep{Anderson04}. Stokes $I$, $Q$ and $U$ images were produced and combined following the guidelines in the ACS Data Handbook \citep{Pavlovsky05}. The optical percentage-polarization and apparent Magnetic Field Position Angle (MFPA) images were computed using $100 \times (Q^2+U^2)^{\frac{1}{2}}/I$ and $\frac{1}{2}\tan^{-1}(U/Q)+90$ degrees, respectively. Corrections were needed for the optical MFPA to include the {\it HST} roll-angle ($285.9$ degrees for this observation) and the offset given by the camera geometry \citep[$-38.2$ degrees for the WFC, from][]{Pavlovsky05}. The Rician bias in the percentage polarization \citep{Serkowski62} was corrected using a script adapted from the Space Telescope European Coordinating Facility (ST-ECF) package \citep{Hook00} in {\sc iraf}, following \citet{Wardle74}. The script uses a `most-probable value' estimator, excluding pixels where the signal-to-noise ratio is $<0.5$, or where the most probable value in the percentage polarization is negative or $>100$ per cent. Small corrections were made for the parallel and perpendicular transmittance of each filter.

Galaxy subtraction was performed using the tasks {\sc ellipse} and {\sc bmodel} in {\sc iraf}, and {\sc imcalc} in {\sc stsdas}. The {\it HST} point-spread function (PSF) was modelled for the ACS/F606W instrument/filter combination using {\sc tinytim} version 6.3 (http://www.stsci.edu/software/tinytim/tinytim.html), and measured to have a FWHM of 0.156 arcsec. The optical images were subsequently read into {\sc aips} for analysis.

\begin{figure*}
\begin{minipage}{160mm}
\centering
\includegraphics[width=160mm]{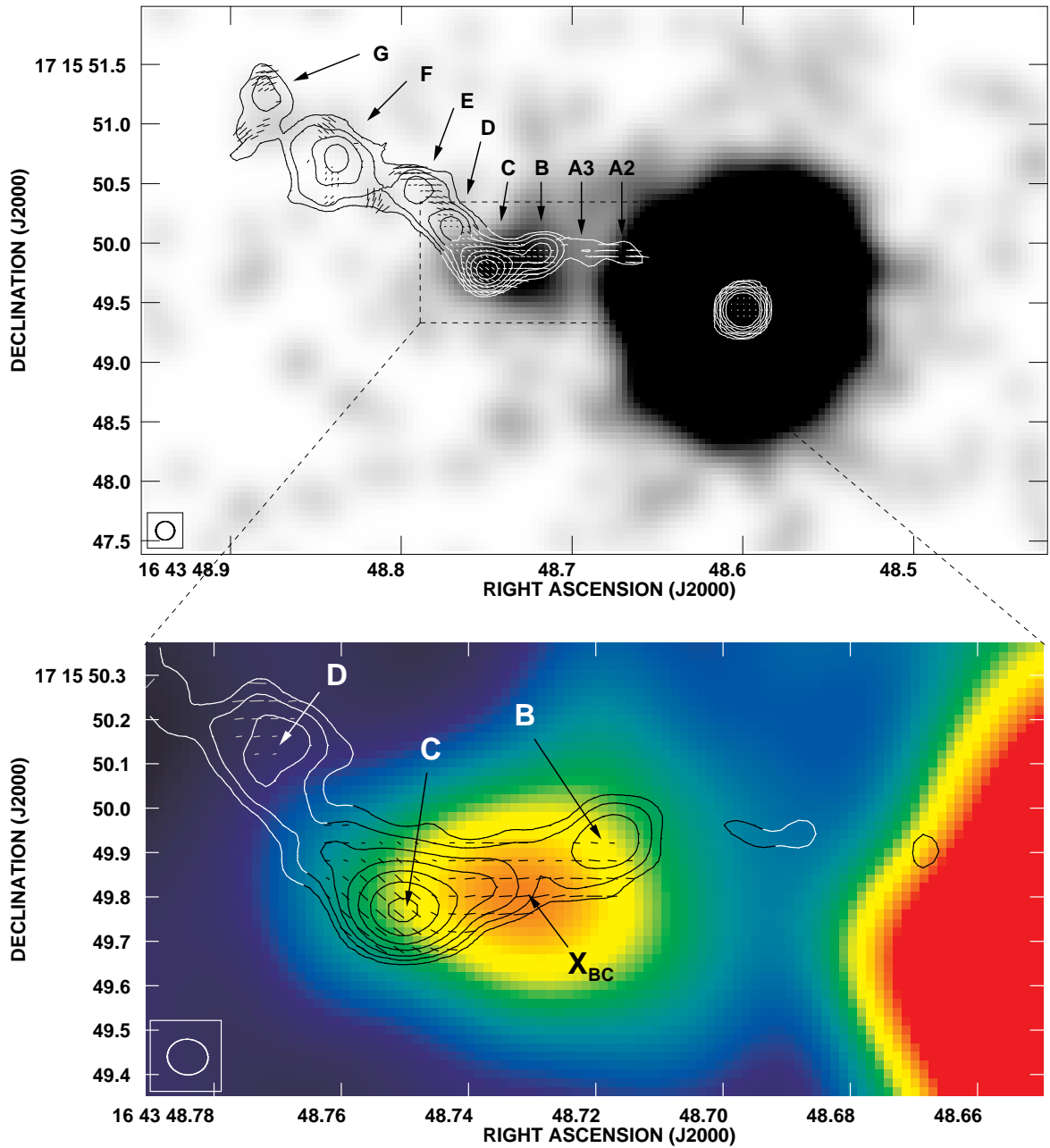}\\
\caption{Top Panel: 22.5~GHz radio intensity contours and polarization vectors (rotated through $90$ degrees), overlaid on a heavily clipped greyscale {\it Chandra} X-ray image smoothed with a 0.35 arcsec (FWHM) circular Gaussian. Contour levels are $0.36~\times~(1, 2, 4, 8, 16, 32)$ mJy beam$^{-1}$, and a vector of length 0.15 arcsec corresponds to 100 per cent polarization. The angular resolution of the radio image is $0.118 \times 0.108$ arcsec. Bottom Panel: Our highest-resolution radio map, created using a restoring beam of $0.092 \times 0.080$ arcsec. The figure shows the region around the brightest jet knot superimposed on a colour scale of the X-ray emission. The offset between the X-ray and radio peak positions is clear.}
\label{fig:xray}
\end{minipage}
\end{figure*}

\begin{table*}
\begin{minipage}{160mm}
\caption{Details of the radio, optical and X-ray observations used in this study of the jet in 3C~346. Integration times are given for the total observation period, before processing.}
\label{tab:obs}
\begin{tabular}{llllll}
\hline
Instrument, Band/Filter & ID & Integration Time (s) & Date & $\nu_{\rm min}$ (Hz) & $\nu_{\rm max}$ (Hz) \\
\hline
{\it VLA} A-config, U-band      & AO127 & $5.49 \times 10^{3}$  & 1996 Dec 14 & $1.49 \times 10^{10}$ & $1.50 \times 10^{10}$ \\
{\it VLA} A-config, K-band      & AO127 & $5.49 \times 10^{3}$  & 1996 Dec 14 & $2.24 \times 10^{10}$ & $2.25 \times 10^{10}$ \\
{\it HST} ACS/WFC1, F606W + POL & 9847  & $1.51 \times 10^{3}$  & 2003 Aug 19 & $4.04 \times 10^{14}$ & $6.76 \times 10^{14}$ \\
{\it Chandra} ACIS, 0.3--6 keV  & 3129  & $44.77 \times 10^{3}$ & 2002 Aug 3  & $7.24 \times 10^{16}$ & $1.45 \times 10^{18}$ \\
\hline
\end{tabular}
\end{minipage}
\end{table*}

\subsection{X-ray data}\label{subsec:xraydata}
The ACIS-S instrument on {\it Chandra} was used to observe 3C~346 on 2002 August 3 (ObsID~3129) in FAINT imaging mode, using chip S3 with a 0.84-second frame time and a 256-row subarray. The level-1 X-ray data were reprocessed to take advantage of more appropriate calibration products. New bad-pixel files were created following the analysis `threads' (online at http://cxc.harvard.edu/ciao/threads/), and a new level-1 event file was created with pixel-randomisation removed. The event file was filtered for good grades with values 0, 2, 3, 4 and 6, and time intervals were excluded where the background rate was larger than 3-sigma from the mean. The mean background rate was 0.3 counts per second, and the exclusion threshold was 0.4 counts per second. The new level-2 event file contained 41129 s of good data, selected from the original 44765 s observation.

The X-ray image was resampled to the same (0.05 arcsec) pixel scale as the optical data, which is much smaller than the native (0.492 arcsec) pixel scale or spatial resolution of {\it Chandra}. The resampled image was then smoothed using a two-dimensional circular Gaussian with FWHM 0.35 arcsec for the figures in this paper. The resulting image resolution of $\sim 0.7$~arcsec FWHM is dominated by the width of the PSF.

\subsection{Radio data}\label{subsec:radiodata}
In order to obtain a high resolution radio map suitable for comparison with the {\it HST} images, we reprocessed archival K-band (22.5~GHz) and U-band (14.9~GHz) {\it VLA} data of 3C~346 using the {\sc aips} package from NRAO. A total of 91.5 minutes of data were obtained at each frequency in the A-configuration over two days, starting 1996 December 14, and the {\it VLA} was used in fast-switching mode at 22.5~GHz to keep track of the rapid phase variations that often affect high-frequency interferometer data.

The data were read into {\sc aips} and bad scans were flagged and removed. Baseline corrections were available from the NRAO logs, so these were applied using the task {\sc clcor}. Flux-density and polarization calibration used 3C~286, and 1658+076 was used for phase referencing during both observations. The calibration procedure was performed in two passes, following the guidelines for reducing high-frequency {\it VLA} data (in appendix~D of the {\it AIPS Cookbook}, available online at http://www.aips.nrao.edu/cook.html). Phase-only corrections were applied in the first pass to minimize decorrelation in the calibrator scans, and only then was the full amplitude-and-phase calibration applied in the second pass to determine the flux density scale. The primary calibrator (3C~286) is resolved by the {\it VLA} at high frequencies and in extended configurations, so both passes used a clean-component model of 3C~286 to represent the small-scale structure on the longer baselines. A good calibration was obtained by tracking the rapid phase variations using a sufficiently short (3 second) interpolation time in the calibration table for the K-band data. The bright core of 3C~346 also provides an excellent anchor for self-calibration, allowing the data to be phase and amplitude self-calibrated to good convergence in Stokes $I$. Stokes $Q$ and $U$ images were then produced from this calibrated data set.

All the radio images were made on a 0.02 arcsec grid. Low-noise images of 3C~346 were produced using the K-band data with near-natural weighting ({\sc robust} = 4), to give an elliptical Gaussian beam of size $0.118 \times 0.108$ arcsec and major axis in position angle 90 degrees. This gives a noise level in Stokes $I$ of $\sim 75$~{$\mu$}Jy beam$^{-1}$. Our highest-resolution K-band image has a beam of $0.092 \times 0.080$ arcsec. The U-band data were imaged with a larger ($0.130 \times 0.120$ arcsec) beam, and have a lower noise level of $\sim 40$~{$\mu$}Jy beam$^{-1}$. The percentage-polarization and apparent MFPA images were produced as described in section~\ref{subsec:opticaldata}. The Rician bias in the percentage polarization \citep{Serkowski62} was corrected by measuring the noise levels in the Stokes $Q$ and $U$ images and using the {\sc polc} algorithm in the {\sc aips} task {\sc comb}. Component fluxes were calculated by fitting two-dimensional Gaussians using the task {\sc imfit} in {\sc aips}.

A radio spectral-index map was produced by combining the 14.9~GHz and 22.5~GHz {\it VLA} data. Both data sets were matched in resolution and carefully aligned before combining them. The 14.9~GHz data was shifted by $(-0.009, 0.016)$ arcsec relative to the 22.5~GHz data to ensure good alignment of the core in each image.

To compare the radio and optical data, the radio images were first regridded to match the 0.05 arcsec pixel scale of the optical images, and then convolved with an elliptical Gaussian to give the radio beam the same FWHM (0.156 arcsec) as the optical PSF.

Since the absolute astrometry of radio images from the {\it VLA} is more accurate than either the optical or X-ray data, all the cores were registered to the coordinates of the radio peak (at $\rmn{RA}= 16^{\rmn{h}} 43^{\rmn{m}} 48\fs600$, $\rmn{Dec.}= +17\degr 15\arcmin 49\farcs44$ in J2000) prior to any analysis. The optical images were shifted by $(-0.17, -0.17)$ arcsec and the X-ray by $(0.05, 0.06)$ arcsec in (RA, Dec).

\section{Results}\label{sec:results}
\subsection{Jet morphology}\label{sec:morphology}

The eastern jet of 3C~346 shows a variety of knots and a dramatic kink in the radio and optical images, shown in Figures~\ref{fig:xray} -- \ref{fig:polarimetry}. The knots are labelled from A--G with increasing distance from the nucleus, following \citet{Perlman06}. Knots A$_{1}$--A$_{3}$ are very faint: knot~A$_{1}$ is not detected in our 22.5~GHz radio image, but we do detect it using the 14.9~GHz data.

\citet{Cotton95} used radio Very Long Baseline Interferometry (VLBI) data to show that the jet is initially straight, from the point at which it is visible against the bright galactic-scale emission until it curves to form the brighter knots B~and~C. At knot~C the jet undergoes an apparent kink of $\sim70$ degrees, although this does not disrupt it: the jet continues to curve and shows additional knots as it moves into the eastern lobe. \citet{deKoff96} and \citet{deVries97} noted the striking similarity between the radio and optical jet morphologies, and our matched {\it HST} and {\it VLA} images of the jet are also in excellent agreement (Figure~\ref{fig:polarimetry}). However, we note that the jet appears to be somewhat more `knotty' in the optical than in the radio: the optical data show that knot~C has a more extended southern edge, and all the knots in the optical image appear to be more distinct than in the smoother radio jet. This phenomenon has been observed in other kiloparsec-scale jets \citep[e.g.][in the case of 3C~273]{Bahcall95}.

\citet{Worrall05} reported the discovery of X-ray synchrotron emission associated with knot~C. With the improved X-ray calibration and radio maps that we are using, we confirm their claim of a positional offset between the X-ray and radio peaks. The X-ray image shows that the centroid of X-ray peak, labelled X$_{\rm BC}$ in Figure~\ref{fig:xray}, is located at $\rmn{RA}= 16^{\rmn{h}} 43^{\rmn{m}} 48\fs731$, $\rmn{Dec.}= +17\degr 15\arcmin 49\farcs80$ (measured using the {\sc zhtools} software), which is 0.29 arcsec away from the peak of the radio emission (at $\rmn{RA}= 16^{\rmn{h}} 43^{\rmn{m}} 48\fs750$, $\rmn{Dec.}= +17\degr 15\arcmin 49\farcs78$). With $\sim70$ X-ray counts in this region, the peak should be located to better than 60~mas using {\it Chandra}, so the offset in peak positions is significant at the $4\sigma$ level.

The integrated 22.5~GHz radio flux density at knot~C was measured to be $54.9 \pm 0.1$ mJy. Its deconvolved size is $0.11 \times 0.07$ arcsec$^2$, corresponding to an ellipsoid with major and minor axes 300~pc and 200~pc, respectively. \citet{Worrall05} measured a 1~keV flux density at knot~C (X$_{\rm BC}$ in Figure~\ref{fig:xray}) of $1.6 \pm 0.2$ nJy, and found the X-ray spectrum consistent with an unabsorbed simple power law of spectral index $\alpha=1.0 \pm 0.3$ (for a spectrum $S_\nu \propto \nu^{-\alpha}$). The radio spectral index variations were mapped using the 14.9~GHz and 22.5~GHz {\it VLA} data. The core shows a spectral index of $\alpha = -0.2 \pm 0.1$, consistent with self-absorbed emission, and the radio spectral index around knot~C shows low-significance variations from $\alpha = 0.4$ to $\alpha = 0.6$.

\subsection{Jet polarimetry}\label{sec:polarimetry}

\begin{figure*}
\begin{minipage}{160mm}
\centering
\includegraphics[width=160mm]{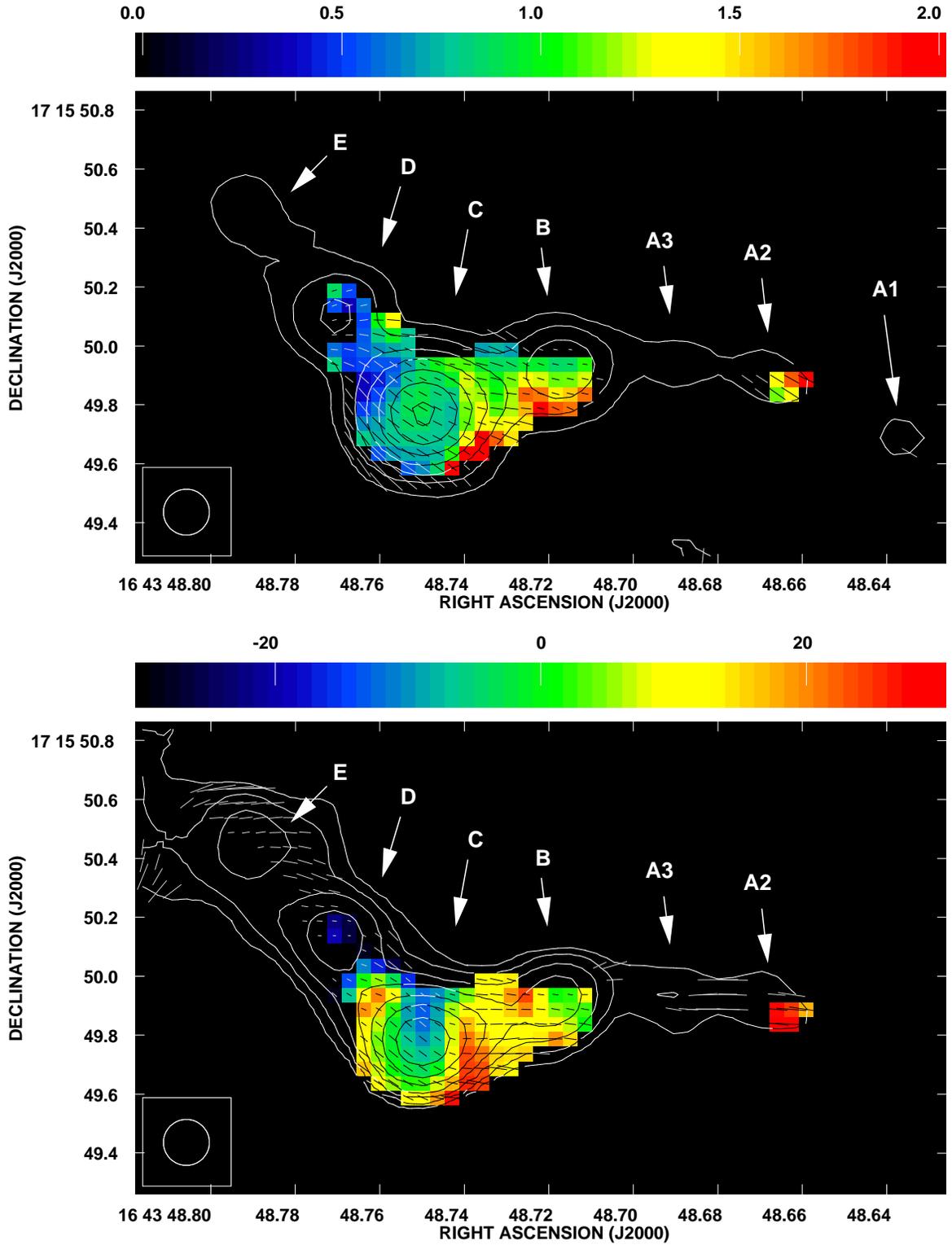}\\
\caption{Top Panel (a): Optical intensity contours and polarization vectors (rotated through 90 degrees). Contour levels are $0.08~\times~(1, 2, 4, 8, 16, 32, 64)$ electrons sec$^{-1}$, and a vector of length 0.15 arcsec corresponds to 100 per cent polarization. The colour scale shows the ratio between percentage of polarized flux detected in the radio and optical bands, where zero indicates relatively little polarized radio flux. Bottom Panel (b): Radio intensity contours and (rotated) polarization vectors to the same angular scale as (a). Contour levels are $0.36~\times~(1, 2, 4, 8, 16, 32)$ mJy beam$^{-1}$, and a vector of length 0.15 arcsec corresponds to 100 per cent polarization. The colour scale shows the difference between the optical and radio magnetic field position angles in degrees.}
\label{fig:polarimetry}
\end{minipage}
\end{figure*}

\begin{table}
\caption{Polarimetry of the knots in the jet of 3C~346 at optical and radio wavelengths, showing the percentage polarization at 1.3~cm ($P_{\rm rad}$), 606~nm ($P_{\rm opt}$), and the difference in apparent magnetic-field position angle (optical-radio), $\Delta_{\chi}$, in degrees.}
\label{tab:polarimetry}
\begin{tabular}{llll}
\hline
Knot & \parbox{1.5cm}{$P_{\rm rad}$, per~cent} & \parbox{1.5cm}{$P_{\rm opt}$, per~cent} & \parbox{2.5cm}{$\Delta_{\chi}$, degrees} \\
\hline
A$_1$ & --          & --         & --           \\
A$_2$ & $65 \pm 10$ & $40 \pm 8$ & $25 \pm 10$  \\
A$_3$ & $65 \pm 10$ & --         & --           \\
B     & $25 \pm 6$  & $18 \pm 8$ & $12 \pm 7$   \\
C     & $20 \pm 4$  & $27 \pm 7$ & $2 \pm 2$    \\
D     & $14 \pm 7$  & $20 \pm 6$ & $-24 \pm 11$ \\
E     & $30 \pm 10$ & --         & --           \\
\hline
\end{tabular}
\end{table}

The structure of the jet's apparent magnetic field is indicated in the optical and radio maps in Figure~\ref{fig:polarimetry}, where the polarization vectors have been rotated through 90 degrees. Although this is common practice, these vectors do not necessarily show the true projected magnetic field position angle (MFPA), which may shift due to relativistic aberration \citep*[e.g.][]{Lyutikov03}.

Error maps of the radio and optical MFPA were produced by measuring the noise on the Stokes $Q$ and $U$ images. In addition to random noise, the optical ACS data suffer from a 3-degree systematic uncertainty in the polarization position angle \citep[see][for further details]{Pavlovsky05}. Polarization vectors are plotted in Figure~\ref{fig:polarimetry} only where the statistical errors in the apparent MFPA are less than 10 degrees: regions where the errors are larger than 10 degrees are shown without vectors and are ignored in this analysis. It should be noted that statistical errors in the apparent MFPA are often much smaller than this, and only about 2 degrees in the brightest knot. The differences in polarization position angle between the 14.9~GHz and 22.5~GHz radio data are small (generally in the range $0-4$ degrees) and, within errors, are consistent with zero. 

Figure~\ref{fig:polarimetry}a (colour-scale) shows the ratio of the percentage polarization detected in the radio and optical bands, where values near zero indicate relatively little polarized radio flux. The data show a weak trend towards increasing optical/radio polarization ratio with distance along the jet: out to knot~C, the jet is relatively more polarized in the radio than in the optical. Near the base, in knots~A$_{2}$ and~A$_{3}$, the radio polarization is high ($65 \pm 10$ per cent) but this falls to $25 \pm 6$ percent in knot~B. In the optical, knot~A$_{2}$ and knot~B are $40 \pm 8$ per cent polarized and $18 \pm 8$ per cent polarized, respectively. No significant optical polarization is detected in A$_{1}$ or A$_{3}$. The radio polarization continues to fall, reaching $20 \pm 4$ per cent at knot~C and $14 \pm 7$ per cent at knot~D. In contrast, the optical polarization does not change much: $27 \pm 7$ per cent at~C, and $20 \pm 6$ per cent at~D. Knots~E and~F are weakly detected in the optical but show no significant polarization, and little optical jet emission is detected after knot~F. 

The radio emission and polarization structure continue after knot~D, however. We note in particular the asymmetric polarization seen across knots~D and~E, which could indicate a helical structure in the magnetic field. The north-western side of knot~E is highly polarized in the radio ($30 \pm 10$ per cent), but the south-eastern side of knot~E shows no significant polarization. The radio polarization is weaker at knot~D (as noted above), but it too shows a similar asymmetry across the jet.

The colour-scale in Figure~\ref{fig:polarimetry}(b) shows differences (in degrees) between the apparent MFPA seen in the optical and radio data. It is clear that the apparent MFPA differences are small throughout the jet and, as shown in Table 2, the optical and radio polarization directions are particularly well aligned in the bulk of knot C. The radio and optical MFPAs in the knot~A complex are broadly parallel to the jet, although they are misaligned by $25~\pm~10$ degrees in knot~A$_{2}$. In knot~B the apparent MFPA remains almost unchanged in the radio, while the apparent optical MFPA rotates through a larger angle to reduce the misalignment: the differences in knot~B are less significant, with a misalignment of $12~\pm~7$ degrees. At knot~C the jet exhibits an apparent 70-degree kink, and both the radio and optical apparent MFPA rotate to remain broadly parallel to the new jet direction. The apparent MFPA differences in the centre of knot~C are consistent with zero, although there is a small region to the south-west of the knot where the differences increase to $\sim25$ degrees and the errors are 8 degrees. On the north-east side of knot~C, the differences are about $-18~\pm~11$ degrees. In knot~D, the misalignment increases to $-24~\pm~11$ degrees. The asymmetric `ridge' of radio polarization mentioned previously (to the north-east of knot~D) does not have a simple polarization structure, although here too the MFPA is broadly parallel to the jet.

\begin{figure}
\centering
\includegraphics[width=80mm]{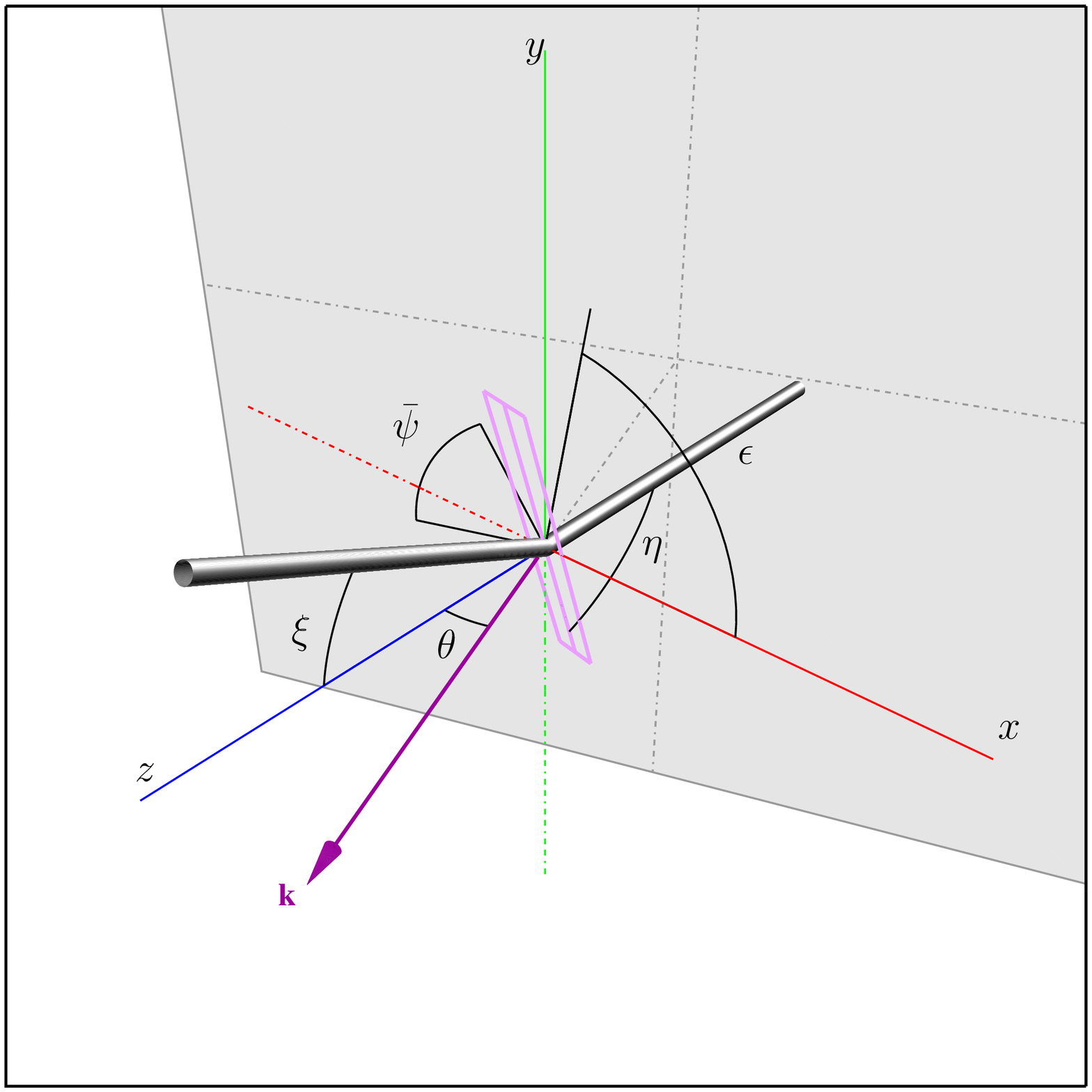}\\
\caption{A three-dimensional perspective view of the geometry used to describe the oblique shock in 3C~346, after \citet{Cawthorne06}. The solid grey area is the plane of the sky for an observer viewing along the $-\vect{k}$ direction, and the plane of the shock contains the origin. The jet (indicated by the shaded cylinder) is directed towards the observer, where the upstream and downstream jet components are split by the shock front. See text for a description of the labelled angles.}
\label{fig:geometry}
\end{figure}

\subsection{Motivation for an oblique shock model}\label{sec:motivation}

\citet{Kirk89} showed that the spectral index of radiation from an oblique shock is flatter than that expected from a parallel shock, which is consistent with our observation of the radio spectral index $\alpha_R$ in the range 0.4--0.6 at knot~C. However, the strongest evidence for the oblique shock can be found in the polarization maps of the jet. The available optical and radio polarimetry (Figure~\ref{fig:polarimetry}) indicates a highly-aligned magnetic field at knot~C, where the jet exhibits a large apparent change of direction. \citet{Worrall05} outlined how such a change could arise if an oblique shock was present at that point, and suggested that such a shock could be caused by the turbulent wake of the companion galaxy to 3C~346 \citep[see figure 9 of][]{Worrall05}. Considering the proposed angle of the shock, they predicted a relatively simple polarization pattern at this point. Using the new polarization measurements, we now go on to explore the properties of such a model and to examine how well it fits our observations.

\section{Modelling the shock}\label{sec:models}

Since the available data on the 3C~346 jet strongly suggests the presence of an oblique shock at knot~C, we have examined this region using models of the polarization from oblique and conical shocks presented by \citet{Cawthorne06}. These models were based on \citet{Cawthorne90} and \citet{Lind85}. The key assumptions of the models are that:

\begin{enumerate}
\item the shock front is a stationary surface;
\item the equation of state ($\Gamma = 4/3$) is that of a relativistic gas, and the magnetic field is dynamically unimportant;
\item the emission is from a thin layer of material just downstream of the shock, so that radiation transfer may be neglected;
\item the shock structure does not evolve over at least one light-crossing time;
\item the emitting material is optically thin;
\item any emission from upstream material is negligible compared to the increase in synchrotron emissivity from material at the shock front.
\end{enumerate}

\begin{figure*}
\begin{minipage}{110mm}
\centering
\includegraphics[width=110mm]{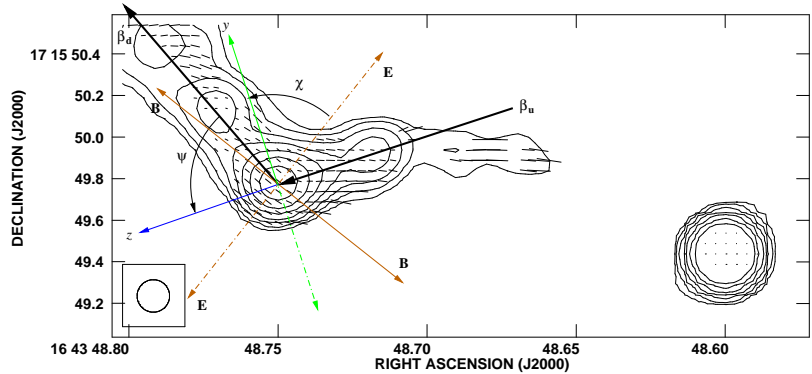}\\
\caption{The geometry of the single-component model as applied to the 3C~346 data, with axes and polarization vectors superposed. The electric-vector position angle is measured with respect to the $y$-axis and is denoted by $\chi$. The electric vector polarization $\mathbf{E}$ is indicated by the dot-dash orange line, and the magnetic vector $\mathbf{B}$ by the solid orange line. This sketch should be used as a guideline only.}
\label{fig:geometry_single_component}
\end{minipage}
\end{figure*}

The geometry is indicated in Figure~\ref{fig:geometry}, where the shock is shown at the origin. The upstream flow direction is shown along the $-z$-axis, and the plane of the shock front is inclined at an angle $\eta$ to the upstream flow. The bulk velocity of the upstream plasma is $\beta_u c$, and the downstream velocity $\beta_d c$ and actual deflection angle of the flow $\xi$ were given by \citet{Lind85} as:

\begin{equation}
\beta_d = \frac{\left[ (1-\beta_u^2 \cos^2\eta)^2 + 9\beta_u^4 \cos^2\eta \sin^2\eta \right] ^{1/2}}{3\beta_u \sin\eta}
\end{equation}

\noindent and

\begin{equation}
\tan\xi = \frac{\tan^2\eta (3 \beta_u^2 - 1) - (1 - \beta_u^2)}{\tan\eta (\tan^2\eta + 1 + 2\beta_u^2)}
\end{equation}

The plane formed by the upstream and downstream flow vectors is rotated by a counterclockwise angle $\epsilon$ relative to the $xz$-plane (when viewed from the $+z$-direction). This structure is viewed by an observer in the plane of the sky along the $-\vect{k}$ direction, which is also in the $xz$-plane and which makes an angle $\theta$ with the $z$-axis. The apparent deflection angle of the flow in the plane of the sky is $\psi$, which is given by

\begin{equation}
\tan\psi = \frac{-\sin\xi \sin\epsilon}{-\sgn(\sin\theta) (\sin\xi \cos\epsilon \cos\theta - \cos\xi \sin\theta)}
\end{equation}

\noindent where the arctangent should be evaluated using all four quadrants, and the angle $\psi$ as defined above is the counterclockwise angle from the projected $z$-axis towards the projected downstream flow direction. Figure~\ref{fig:geometry} shows $\bar\psi$, the 360-degree complement to $\psi$.

In the following sections we examine two models from \citet{Cawthorne06} and apply them to the 3C~346 data to see how the magnetic field geometry may give rise to the observed polarization at the shock. We use the apparent deflection angle $\psi$ as a constraint to examine the distribution of the free parameters in the model: the obliquity of the shock (angle $\eta$), the viewing angles $\epsilon$ and $\theta$, and the upstream flow speed $\beta_u$.

\subsection{Single-component field}\label{sec:singlecomp}
The first model \citep[presented by][]{Cawthorne90} assumes that there is no ordered magnetic field present upstream of the shock (i.e. any upstream field is completely tangled and any emissions from upstream material are unpolarized). The polarization at the shock itself would then arise entirely from the tangled upstream field becoming partially ordered due to compression by the shock. In the 3C~346 jet, significant polarization {\it is} observed upstream of knot~C, but we have nevertheless examined this model to see how well it might reproduce the observed downstream polarization properties. The polarization angle $\chi_t$ of synchrotron radiation from the shock-compressed tangled-field component was given by \citet{Cawthorne06} as

\begin{equation}
\tan\chi_t = \frac{\cos\epsilon \cos\eta (\beta_u - \cos\theta) - \sin\eta \sin\theta}{\sin\epsilon \cos\eta (1 - \beta_u \cos\theta)}
\end{equation}

\noindent where $\chi_t$ is the electric-vector position angle (EVPA), measured counterclockwise from the $y$-axis in the plane of the sky. The fractional polarization of this shocked-tangled component \citep{Cawthorne06} is

\begin{equation}
m_t = m_0 \frac{(1 - \kappa^2) \sin^2\psi'}{2 - \sin^2\psi' (1 - \kappa^2)}
\end{equation}

\noindent where $m_0 = (\alpha + 1)/(\alpha + 5/3)$ is the fractional polarization from a uniform field, which depends on the spectral index $\alpha$ of optically thin synchrotron emission. The compression coefficient ($\kappa$, the ratio of the upstream to downstream number densities) for an upstream Lorentz factor $\gamma_u$ was given by \citet{Cawthorne90} as

\begin{equation}
\kappa = \frac{(1 - \beta_u^2 \cos^2\eta)^{1/2}}{\gamma_u \beta_u \sin\eta (8\beta_u^2 \sin^2\eta - \gamma_u^{-2})^{1/2}}.
\end{equation}

The quantity $\psi'$ is the angle between the shock normal and the line-of-sight in the rest frame of the downstream plasma, and is found by resolving $\vect{k}$ into components parallel and perpendicular to the upstream flow in its own rest frame \citep[$k_{\parallel}''$ and $k_{\perp}''$ respectively, from][]{Cawthorne06}: using these, the component normal to the shock ($k_n''$) can be found and transformed into the rest frame of the downstream plasma to give an expression for $\psi'$. The component normal to the shock is $k_n'' = k_{\parallel}'' \sin\eta'' - k_{\perp}'' \cos\eta''$, where $\tan\eta'' = \gamma_u \tan\eta$, $k_{\parallel}'' = (\cos\theta - \beta_u)/(1 - \beta_u \cos\theta)$ and $k_{\perp}'' = \sin\theta \cos\epsilon / (\gamma_u[1 - \beta_u \cos\theta])$ \citep{Cawthorne06}. In the rest frame of the upstream flow, the downstream velocity is expressed as $\beta_d'' = 3\beta_u \sin\eta''/2 - 1/(2\beta_u \sin\eta'')$, and so finally

\begin{equation}
\label{eqn:psi_prime}
\cos\psi' = \frac{k_n'' + \beta_d''}{1 + k_n''\beta_d''}.
\end{equation}

\begin{figure}
\centering
\includegraphics[width=80mm]{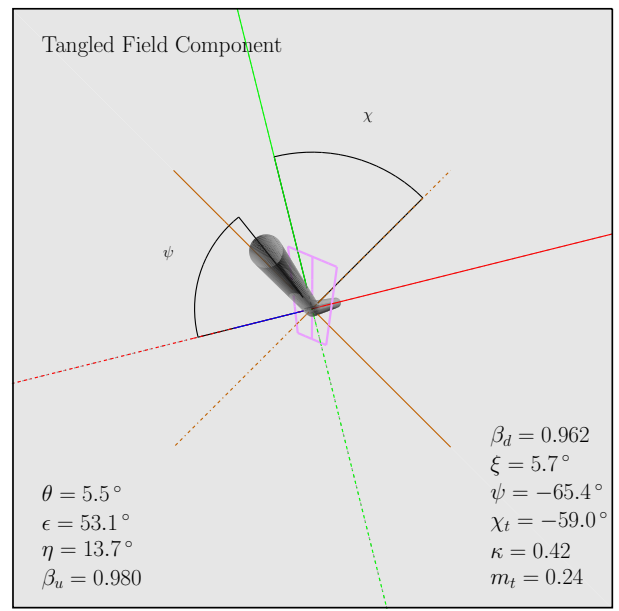}\\
\caption{An example of the solutions allowed using the single-component model with only tangled field. The figure shows the model viewed in the plane of the sky, with $\vect{k}$ pointing out of the plane of the paper (cf. Figure~\ref{fig:geometry}).  Although this model can reproduce our observations (Figure~\ref{fig:geometry_single_component}), the geometry of these solutions at such a small angle to the line-of-sight is implausible: see text for discussion.}
\label{fig:singlePolResults}
\end{figure}

\subsection{Results using the single-component model}\label{sec:results1}

We search for sets of ($\beta_u$, $\sin\theta$, $\sin\eta$, $\cos\epsilon$) parameters that can reproduce the jet kink and polarization. From the radio data, the measured spectral index $\alpha$ is $0.55 \pm 0.05$ at the shock (knot~C; see Section~\ref{sec:morphology}), and the apparent deflection angle of the radio jet (as defined in Section~\ref{sec:models}) is $-67 \pm 1$ degrees. The observed EVPA at the shock is $-57 \pm 2$ degrees, and the polarization fraction $m_t = 0.25 \pm 0.02$ (see Figure~\ref{fig:geometry_single_component}). With these constraints, we step in $\beta_u$ using a step-size of 0.01 from 0.60--0.99, and we map the parameter space with a resolution of 0.005 in the range $0 \le \sin\theta \le 1$, $0 \le \sin\eta \le 1$ and $-1 \le \cos\epsilon \le 1$. Sets of ($\beta_u$, $\sin\theta$, $\sin\eta$, $\cos\epsilon$) parameters that can reproduce the observed polarization features are stored as possible solutions -- otherwise, they are rejected.

The only solutions allowed by our data for this model lie at small angles to the line-of-sight ($\theta < 6$ degrees), where the obliquity of the shock is large ($\eta < 15$ degrees) and the upstream plasma is highly relativistic ($\beta_u > 0.98$). The allowed parameter ranges are given in the first section of Table~\ref{tab:results}, and an example is shown in Figure~\ref{fig:singlePolResults}. Although the essential features of the shock can be reproduced using this model, these solutions lead to implausible geometries: such a small angle to the line-of-sight would mean that the nearest upstream radio knot (knot~B) is at least 15~kpc from the shock. Also, because 3C~346 is an FR~II radio source, the core should appear more like a quasar than a radio galaxy, with broad emission lines, if the jet follows a straight path at this angle all the way back to the nucleus. This is inconsistent with observations \citep[see][]{Worrall01a}. \citet{Worrall05} also showed that an angle to the line-of-sight $\theta \sim 20$ degrees was compatible with the large-scale bending of the jet due to pressure gradients in the cluster gas. Since we believe the cluster atmosphere produces the large-scale gradual bending of the flow from the core out to knot~E, we reject the single-component model and instead investigate improved models which take into account the upstream magnetic field.

\begin{table}
\caption{The results of modelling the jet in 3C~346 using the models of \citet{Cawthorne06}. The minimum, best-fit and maximum values are shown for each parameter. The best-fit values and errors were estimated using the median, lower quartile and upper quartile of the parameter distribution. See text for discussion.}
\begin{tabular}{c|ccc}
\label{tab:results}
Parameter      & Min                    & Best-fit value                           & Max     \\
\hline
\multicolumn{4}{l}{Single-component model: Tangled upstream field}                           \\[6pt]
$\beta_u$      & 0.98                   & $0.99_{-0.01}^{+0.00}$                   & 0.99    \\[6pt]
$\sin\theta$   & 0.055                  & $0.065_{-0.000}^{+0.025}$                & 0.100   \\[6pt]
$\sin\eta$     & 0.150                  & $0.165_{-0.000}^{+0.065}$                & 0.245   \\[6pt]
$\cos\epsilon$ & 0.570                  & $0.605_{-0.015}^{+0.015}$                & 0.675   \\[6pt]
\hline
\multicolumn{4}{l}{Two-component model: Uniform and tangled upstream field}                  \\[6pt]
$\beta_u$      & 0.60                   & $0.91_{-0.07}^{+0.05}$                   & 0.99    \\[6pt]
$\sin\theta$   & 0.005                  & $0.240_{-0.115}^{+0.135}$                & 0.655   \\[6pt]
$\sin\eta$     & 0.105                  & $0.775_{-0.135}^{+0.105}$                & 0.995   \\[6pt]
$\cos\epsilon$ & -0.660                 & $0.070_{-0.335}^{+0.360}$                & 0.995   \\[6pt]
$f$            & 0.21                   & $0.32_{-0.09}^{+0.24}$                   & 1.00    \\[6pt]
$g$            & 0.00                   & $0.12_{-0.05}^{+0.09}$                   & 0.98    \\[6pt]
$I_t$ (mJy)    & 1.0                    & $47.8_{-4.7}^{+2.8}$                     & 54.3    \\[6pt]
$Q_t$ (mJy)    & 4.4                    & $5.1_{-0.3}^{+0.5}$                      & 9.8     \\[6pt]
$U_t$ (mJy)    & -9.5                   & $-4.9_{-2.0}^{+3.3}$                     & 28.4    \\[6pt]
$I_u$ (mJy)    & 0.002                  & $6.5_{-2.8}^{+4.7}$                      & 53.3    \\[6pt]
$Q_u$ (mJy)    & -5.3                   & $-0.7_{-0.5}^{+0.3}$                     & -0.0002 \\[6pt]
$U_u$ (mJy)    & -38.0                  & $-4.6_{-3.3}^{+2.0}$                     & -0.001  \\[6pt]
\end{tabular}
\end{table}

\subsection{Two-component field}\label{sec:twocomp}

The improved model treats the magnetic field upstream of the shock as two components. As in Section~\ref{sec:singlecomp}, one component of the upstream field is completely tangled, and contributes to the observed polarization only after becoming ordered at the shock-front. The remainder of the magnetic energy is present in a uniform field parallel to the upstream flow: since this field is frozen into the plasma, the uniform-field component will remain parallel to the jet in the downstream region \citep{Cawthorne06}. 

The electric vector polarization position angle of the uniform-field component, $\chi_u$, at the shock is given by

\begin{equation}
\label{eqn:chi_u}
\tan\chi_u = \frac{\sin\xi \sin\epsilon}{\sin\xi \cos\epsilon \cos\theta - \cos\xi \sin\theta},
\end{equation}

\noindent and since the field is uniform, the fractional polarization of this component will simply be $m_0$.

We examine the distributions of Stokes parameters for the shocked-tangled-component ($I_t$, $Q_t$ and $U_t$) and uniform-field components ($I_u$, $Q_u$ and $U_u$) in addition to the distributions of geometric parameters ($\theta$, $\eta$, $\epsilon$) and the speed of the upstream flow ($\beta_u$). Using this model, we assume that all the emission from the bright knot originates from a small region immediately downstream of the shock front. In this post-shock region, the observable Stokes parameters are denoted $I_o$, $Q_o$ and $U_o$ (Stokes $V$ is zero), and the downstream spectral index is $\alpha_o$. The Stokes parameters are related by

\begin{equation}\label{Eqn:StokesVector}
\left( \begin{array}{ccc}
I_o \\
Q_o \\
U_o \end{array} \right)
=
\left( \begin{array}{ccc}
I_u \\
Q_u \\
U_u \end{array} \right)
+
\left( \begin{array}{ccc}
I_t \\
Q_t \\
U_t \end{array} \right)
\end{equation}

The intensity due to the shocked uniform-field component is given by

\begin{equation}
I_u = g I_o,
\end{equation}

\noindent where the ratio $g$ is derived in Section~\ref{sec:g}. Since we infer the value of $\chi_u$ from the jet direction (Equation~\ref{eqn:chi_u}), we then use

\begin{equation}
\label{eqn:Q_u}
Q_u = \pm \frac{I_u m_0}{\sqrt{1 + \tan^{2} 2\chi_u}}
\end{equation}

\noindent and 

\begin{equation}
U_u = Q_u \tan 2\chi_u
\end{equation}

\noindent to find the polarization from the uniform-field component in the post-shock region. The polarization from the shocked-tangled component was then obtained using Equation~\ref{Eqn:StokesVector}.

\subsubsection{The fractional intensity of the uniform-field component}\label{sec:g}

The fraction of Stokes $I$ from the uniform-field component in the downstream (post-shock) region is defined using

\begin{equation}
g = \frac{I_u}{I_o} = \frac{I_u}{I_u + I_t}.
\end{equation}

Using equation~10 of \citet{Cawthorne06},

\begin{equation}
I_u \propto \left(B_{p,d} \sin\Omega'\right)^{1 + \alpha_o},
\end{equation}

\noindent where $\alpha_o$ is the observed post-shock spectral index, $B_{p,d}$ is the downstream magnetic field parallel to the jet, and $\Omega'$ is the angle between the line of sight and magnetic field in the rest frame of the emitting gas. The angle between the downstream flow and line of sight in the shock frame is $\Omega$ \citep{Cawthorne06}, given by

\begin{equation}
\cos \Omega = \sin\theta \sin\xi \cos\epsilon + \cos\theta \cos\xi,
\end{equation}

\noindent which may be transformed into the rest frame of the downstream flow to find $\Omega'$ using

\begin{equation}
\cos \Omega' = \frac{\cos\Omega - \beta_d}{1 - \beta_d \cos\Omega}.
\end{equation}

The magnetic field $B_{p,d}$ can be found using equation~8 of \citet{Cawthorne06}:

\begin{equation}
\label{eqn:B_p}
B_{p,d} = B_{p} \left[\frac{1 + \kappa^2 \gamma_u^2 \tan^2 \eta}{\kappa^2 \left(1 + \gamma_u^2 \tan^2 \eta \right)}\right]^{\frac{1}{2}} = h B_{p}
\end{equation}

\noindent where $h$ has been introduced for convenience. Using equation~14 from \citet{Cawthorne06}, we assume that 

\begin{equation}
\label{eqn:I_t}
I_t \propto \left[\frac{B_t^{2}}{3 \kappa^{2}}\left(2 - \sin^{2}\psi'\left(1 - \kappa^2\right) \right)\right]^{\frac{1 + \alpha_o}{2}}
\end{equation}

\noindent (with $\psi'$ given by Equation~\ref{eqn:psi_prime}) will be a reasonable approximation for $I_t$ for values of $\alpha_o$ close to $1$. In Equations~\ref{eqn:B_p} and~\ref{eqn:I_t}, $B_p$ and $B_t$ are the parallel and tangled components of {\it upstream} magnetic field, respectively. Introducing $f$ as the fraction of energy in ordered upstream field (see Section~\ref{sec:f}),

\begin{equation}
B_{p}^{2} = f \left(B_{p}^{2} + B_{t}^{2}\right),
\end{equation}

\noindent and 

\begin{equation}
B_{t}^{2} = (1 - f) \left(B_{p}^{2} + B_{t}^{2}\right),
\end{equation}

\noindent by definition. After some algebra, we obtain

\begin{equation}
g = \frac{1}{1 + \left[\frac{1 - f}{f} \cdot \frac{2 - \sin^2 \psi' (1 - \kappa^2)}{3 \kappa^2 h^2 \sin^2 \Omega'}\right]^{\frac{1 + \alpha_o}{2}}}
\end{equation}

\subsubsection{The fraction of energy in ordered magnetic field}\label{sec:f}

Following similar methods to those in Section~\ref{sec:g}, the fraction of energy in ordered magnetic field is found using the upstream polarization fraction $P_i$ and spectral index $\alpha_i$:

\begin{equation}
f = \frac{1}{1 + \sin^{2}\theta'\left[\frac{1}{s_{\alpha}}\left(\frac{m_{0}}{P_{i}} - 1\right)\right]^{\frac{2}{1 + \alpha_i}}}
\end{equation}

\noindent where $\theta'$ is the angle between the line of sight and the magnetic field in the rest frame of the emitting gas, which is obtained using

\begin{equation}
\cos \theta' = \frac{\cos\theta - \beta_u}{1 - \beta_u \cos\theta}.
\end{equation}

The parameter $s_{\alpha}$ depends on $\alpha_i$ according to

\begin{equation}
s_{\alpha} = 2^{1 + \alpha} \frac{\left[\Gamma\left(\frac{3}{2} + \frac{1}{2}\alpha\right)\right]^{2}}{\Gamma\left(3 + \alpha\right)},
\end{equation}

\noindent where $\Gamma$ is the Gamma function.

\subsection{Results using the two-component model}\label{sec:results2}

To apply the two-component model to the oblique shock in 3C~346, we treat the upstream flow as travelling towards knot~C from the centre of the X-ray-emitting region (knot~X$_{\rm BC}$), rather than from radio knot~B: this ensures the upstream magnetic field position angle (MFPA) is parallel to the upstream flow (see Figure~\ref{fig:xray}), which is the key assumption that allows us to consider this model. The upstream polarization fraction is $P_i = 0.22 \pm 0.02$, and spectral index $\alpha_i = 0.7 \pm 0.1$. Downstream, the spectral index is $\alpha_o = 0.55 \pm 0.05$, and the integrated Stokes parameters are $I_o = 54 \pm 0.091$~mJy, $Q_o = 4.4 \pm 0.2$~mJy and $U_o = -9.5 \pm 0.1$~mJy. The apparent deflection angle is now $\psi = -50 \pm 2$ degrees, and with respect to the new $y$-axis, we infer the EVPA of the uniform-field component to be $\chi_u = -49 \pm 3$ degrees. We therefore choose the negative root for $Q_u$ in Equation~\ref{eqn:Q_u}.

As in Section~\ref{sec:results1}, sets of ($\beta_u$, $\sin\theta$, $\sin\eta$, $\cos\epsilon$) parameters that can reproduce the observed polarization features are stored as possible solutions. The results of mapping the parameter space of the two-component model within the ranges given in Section~\ref{sec:results1} are shown in the second section of Table~\ref{tab:results}. 

The distribution of $\beta_u$ is now very wide, ranging from $\beta_u = 0.60$ to $\beta_u = 0.99$ -- no upper limit to $\beta_u$ was found. The median value indicates that the upstream flow is highly relativistic, with $\beta_u = 0.91_{-0.07}^{+0.05}$. The maximum allowed value for the angle to the line of sight ($\theta$) is 41 degrees, and while the distribution is not strongly peaked, the median value is significantly lower at $\theta = 14_{-7}^{+8}$ degrees. The distribution of the obliquity angle $\eta$ has a sharper peak with a modal value close to $\eta \simeq 64$ degrees, although the median is at $\eta = 51 \pm 11$ degrees. The viewing angle $\epsilon$ is not well-constrained, since it has a distribution with a large spread around $\epsilon = 86_{-21}^{+19}$ degrees, and a maximum allowed value near 130 degrees. The fraction of energy upstream in ordered magnetic field is $f = 0.319_{-0.086}^{+0.242}$, and ranges from 0.2 to 1.0, although the distribution of $g$ (the fractional intensity from the uniform-field component) is peaked strongly around low values, such that $g = 0.120_{-0.052}^{+0.086}$. A valid example solution that lies within all constraints and errors has $\beta_u = 0.92$, $\theta = 20$ degrees, $\eta = 55$ degrees and $\epsilon = 95$ degrees. Other solutions exist close to the median values given in Table~\ref{tab:results}.

\section{Discussion}

We have shown that the data support the presence of an oblique shock in the jet of 3C~346 (see Section~\ref{sec:motivation}). The first model we applied to the shock required an unrealistic geometry to fit our data. The improved model of \citet{Cawthorne06} produced better results, although our simulations were not able to provide unambiguous solutions to the geometry of the shock in 3C~346. Our results suggest that the jet (before the shock) is being viewed at an angle of roughly 20 degrees to the line of sight, and that the jet speed before the shock is highly relativistic, with likely values close to $\beta_u = 0.91_{-0.07}^{+0.05}$.

The models described in the preceding sections make important assumptions about the state of the shock, as listed in Section~\ref{sec:models}. The first of these, the assumption that the shock front is a static surface, means that we are unable to use the X-ray emission lifetime to place any additional constraints on the shock geometry. Since the X-ray emission is localised in a region which is unresolved using {\it Chandra}, it would be possible, in principle, to use the X-ray emission lifetime to constrain the geometry: this would require the X-ray emitting plasma at the shock front to be moving at an apparent velocity such that the time taken to traverse an angular distance of order the size of the {\it Chandra} PSF is greater than the X-ray emission lifetime in the rest frame of the observer. \citet{Worrall05} reported synchrotron emission from the X-ray knot, where the best-fitting X-ray lifetime is $k \delta^2$ for a given Doppler-beaming factor $\delta$, and $k = 4.5$ years. Assuming a maximum linear size for the X-ray emission $a$, the required condition is

\begin{equation}
\frac{a}{kc} \ge \frac{\beta_s \sin\theta (1-\beta_s^2)}{(1-\beta_s \cos\theta)^3},
\end{equation}

\noindent where $\beta_s$ is the shock speed. However, because the shock is assumed to be static ($\beta_s = 0$), this argument cannot be used. We note that the peak of the X-ray emission shows a significant offset to the radio and optical peak positions, with a separation $0.29 \pm 0.06$ arcsec ($0.80 \pm 0.17$~kpc). Similar offsets have been observed in other kpc-scale jets \citep[e.g.][in 3C~15 and 3C~66B, respectively]{Dulwich07,Hardcastle01}, although their underlying cause is currently not well-understood. Even if the shock speed was included in the model and constrained using the X-ray emission lifetime, the geometry would remain under-determined, since there are still more free parameters than observables. Another possibility is that knot~B and~C may both represent a complex of knots -- unresolved at the redshift of 3C~346 -- any one of which could be the X-ray emitter.

The other key assumption of the model is that the ordered component of magnetic field upstream of the shock is parallel to the upstream flow, and remains parallel to the downstream flow after passing through the shock. For the MFPA to be parallel to the upstream flow in 3C~346, we must assume that $\beta_u$ is directed at radio knot~C from the upstream X-ray knot, rather than from radio knot~B. It seems reasonable that radio knot~C and the X-ray emission form part of the same structure, and since the X-ray peak falls between knots~B and~C in our highest-resolution radio image, we feel justified in using the X-ray knot to define the upstream flow direction relative to the shock at knot~C. The upstream radio jet also shows significant curvature from the core through knots~A$_{1}$--A$_{3}$ and~B, so we cannot determine a conclusive $\beta_u$ direction using the radio data alone. On the downstream side, radio knots~C, D and E show a good linear alignment, so we can determine the direction of $\beta_d$ with some confidence.

The MFPA upstream of knot~C is also open to interpretation, depending on whether the optical or radio data are used. As shown in Figure~\ref{fig:polarimetry}, the region between knots~B and~C shows the biggest differences in the optical and radio polarization position angles, which can be as large as $\sim 20$ degrees. At the highest angular resolution, our 0.08-arcsec, 22.5~GHz radio polarimetry shows the apparent field to be parallel to the upstream flow in the region where the X-ray emission is brightest, although the optical data show a significant difference in polarization angle of $10-15$ degrees. \citet{Dulwich07} showed that such differences may arise due to a stratified jet, where different components moving at different speeds and emitting in different wavebands could give rise to apparent differences in the polarization position angle of the radiation from each component, even if the magnetic field in each is intrinsically the same. A similar phenomenon could be occurring upstream of the strong shock in 3C~346, if velocity gradients across the jet are large and give rise to a spine-sheath structure in this region. This is consistent with our observation that the fractional polarization in the optical is less than that seen in the higher-resolution radio data between knots~B and~C, which may result if the optical emission originates on smaller scales and is more prone to the effects of beam-averaging.

The polarization position angles at knot~C are well-aligned in our radio and optical data, so we have used the MFPA at this point to determine the apparent magnetic field at the shock front. The most likely geometry of the shock, as determined by our modelling in Section~\ref{sec:results2}, is in good agreement with the jet geometry adopted by \citet{Worrall05}. Our results agree with \citet{Giovannini01} and \citet{Cotton95}, who argued that the angle to the line of sight ($\theta$) should be less than about $30$ degrees, based on jet-sidedness and the observed properties of the core. \citet{Worrall05} also analysed the galaxy-gas and cluster-gas properties using the X-ray radial profile, and observed that the pressures of the two gas components would be equal at the position of the X-ray bright knot if $\theta = 18^{+5}_{-12}$ degrees, which is broadly consistent with our median value ($\theta = 14_{-7}^{+8}$ degrees).

\section{Summary}\label{sec:summary}

We have presented multi-band observations of the jet in 3C~346 using high-resolution data from the {\it VLA}, {\it HST} and {\it Chandra} observatories. The jet has similar optical and radio morphologies, while the X-ray emission from the shocked region is significantly offset from the optical and radio peak positions. The optical and radio polarimetry show close correspondence in the apparent magnetic field position angle and the fractional polarization at knot~C, where the jet undergoes a large apparent kink of almost 70 degrees in the optical and radio images, and which we interpret as an oblique shock. We have explored models of the polarization from oblique shocks presented by \citet{Cawthorne06} to understand the geometry of the 3C~346 jet, and we justify the key assumptions required to apply the models to our data. Although our modelling has not provided a unique description of the jet in 3C~346, we find that the upstream flow is likely to be moving with velocity $\beta_u = 0.91_{-0.07}^{+0.05}$, where the plane of the shock front is inclined at an angle of $\eta = 51 \pm 11$ degrees to the upstream flow which is at an angle $\theta = 14_{-7}^{+8}$ degrees to our line of sight. It has been necessary to include the polarization observed upstream of the shock and model the initial magnetic field using tangled- and uniform-field components, since models that treat the upstream field as completely tangled fail to give a reasonable description of the jet.

\section*{Acknowledgments}
FD acknowledges a research studentship from the UK Science and Technology Facilities Council (STFC), and thanks T.V. Cawthorne for helpful discussions and critical corrections. We also thank the anonymous referee for helpful suggestions to improve the manuscript. Work at UMBC was supported by NASA LTSA grant NNG05-GD63DG and {\it HST} guest observer grant GO-09847.01. The National Radio Astronomy Observatory is a facility of the National Science Foundation operated under cooperative agreement by Associated Universities, Inc. This research has used observations made with the NASA/ESA Hubble Space Telescope, obtained from the data archive at the Space Telescope Institute. STScI is operated by the association of Universities for Research in Astronomy, Inc. under the NASA contract NAS 5-26555. We thank the CXC for its support of {\it Chandra} observations, calibrations, and data processing.

\label{lastpage}

\end{document}